\documentclass[prb,onecolumn,showpacs,floatfix,amsmath,amssymb,superscriptaddress]{revtex4-2}
\usepackage{amsfonts}
\usepackage{stmaryrd}
\usepackage{bbm}
\usepackage{mathrsfs}
\usepackage{tipa}
\usepackage{amssymb}
\usepackage{txfonts}
\usepackage{graphicx}
\usepackage{dcolumn}
\usepackage{epstopdf}
\usepackage[colorlinks,linkcolor=black,urlcolor=black,citecolor=black]{hyperref}
\usepackage{multirow}
\usepackage{subfigure}
\usepackage{url}
\usepackage{upgreek}
\usepackage[utf8]{inputenc}
\usepackage[english]{babel}
\usepackage[locale = US]{siunitx}
\usepackage{lipsum}
\usepackage[normalem]{ulem} 
\usepackage{soul}

\begin{document}
\title{THz Control of Exchange Mode in a Ferrimagnetic Cavity}

\author{C. Reinhoffer}
\affiliation{Institute of Physics II, University of Cologne, 50937 Cologne, Germany}

\author{I. Razdolski}
\affiliation{Faculty of Physics, University of Bialystok, Ciolkowskiego 1L, 15-245 Bialystok, Poland}

\author{P. Stein}
\affiliation{Institute of Physics II, University of Cologne, 50937 Cologne, Germany}

\author{S. Germanskiy}
\affiliation{Institute of Physics II, University of Cologne, 50937 Cologne, Germany}

\author{A. Stupakiewicz}
\affiliation{Faculty of Physics, University of Bialystok, Ciolkowskiego 1L, 15-245 Bialystok, Poland}

\author{P. H. M. van Loosdrecht}
\affiliation{Institute of Physics II, University of Cologne, 50937 Cologne, Germany}

\author{E. A. Mashkovich}
\affiliation{Institute of Physics II, University of Cologne, 50937 Cologne, Germany}

\date{\today}

\begin{abstract}
The interaction of terahertz (THz) radiation with high-frequency spin resonances in complex magnetic materials is central for modern ultrafast magnonics. Here we demonstrate strong variations of the excitation efficiency of the sub-THz exchange magnon in a single crystal ferrimagnet (Gd,Bi)$_3$Fe$_5$O$_{12}$. 
An enhancement of the exchange magnon amplitude is observed when its frequency matches an eigenmode of the cavity created by the sample interfaces.
Moreover, this enhancement is accompanied by a 5-fold decrease in effective damping of the exchange mode.
The THz-exchange magnon interaction in the cavity is analyzed within the developed Landau-Lifshitz-Gilbert formalism for three coupled magnetization sublattices and cavity-enhanced THz field.
This work presents a novel approach for the THz excitation of spin dynamics in ferrimagnets and outlines promising pathways for the controlled optimization of light-spin coupling in single crystals.
\end{abstract}

\maketitle

\section{INTRODUCTION}
In an era where large language models are becoming increasingly dominant in society, the need for faster computational power is more pronounced than ever before~\cite{ajani_advancements_2024}. However, the predominant form of data storage technology today is still ferromagnetic-based, which limits operational frequencies to a few GHz~\cite{markov_limits_2014}. To achieve significantly higher frequencies, i.e. in the THz range, antiferromagnetic magnetic recording and control is the subject of intense investigation nowadays
~\cite{wadley_electrical_2016,pirro_advances_2021,li_light-controlled_2022, grishunin_excitation_2021, qin_natcomm_2021, song_how_2018, mak_probing_2019, sriram_light-induced_2022, parchenko_non-thermal_2016, parchenko_wide_2013, mashkovich_terahertz-driven_2022, metzger_impulsive_2023}. Among media with antiferromagnetic ordering, ferrimagnets are especially appealing as they combine the feasibility of magnetic control with high THz-scale eigenfrequencies. One material class at the center of these investigations has been the rare earth iron garnets~\cite{parchenko_non-thermal_2016, parchenko_wide_2013, mashkovich_terahertz-driven_2022, hsu_observation_2020}. Iron garnets crystallize in the l$\bar{\text{a}}$3d structure where the iron ions form two sublattices sitting in oxygen octahedral $\vec{M}_{\text{a}}$ and tetrahedral sites $\vec{M}_{\text{d}}$ with a ratio of 10~$\mu_\text{B}$:15~$\mu_\text{B}$, while the rare earth ions occupy the interstitial sites. By substituting the main rare earth element \cite{e_p_wohlfarth_ferromagnetic_1999}, and by doping with non-magnetic ions \cite{hansen_saturation_1974}, the magnetic properties of these materials become highly tunable. Due to the strong Fe-Fe exchange interaction, iron garnets typically are treated as ferrimagnets characterized by two sublattices: Gadolinium $\vec{M}_{\text{Gd}} = \vec{M}_{\text{c}}$ and joint iron $\vec{M}_{\text{Fe}}=\vec{M}_{\text{d}}+\vec{M}_{\text{a}}$, respectively. The strength of the exchange interaction between these sublattices can be directly probed by measuring the corresponding magnetic exchange mode~\cite{kaplan_exchange_1953}. The frequency of this resonance is highly sensitive to temperature, especially in the vicinity of the magnetic compensation point, where $\lvert\vec{M}_{\text{Fe}}\rvert$ = $\lvert\vec{M}_{\text{Gd}}\rvert$. Furthermore, substitution of Gd with Bi enhances the magneto-optical response, enabling ultrafast magneto-optical spectroscopy \cite{hansen_magnetic_1984, takeuchi_faraday_1975}. 
The exchange mode can be excited either resonatly~\cite{blank_thz-scale_2021} or via opto-magnetic effects~\cite{parchenko_non-thermal_2016}.

There are several approaches to increase the coupling strength between photons and magnons: some are based on energy concentration using cavities~\cite{zhang_strongly_2014, jarc_tunable_2022, blank_magneto-optical_2023}, while others exploit multiple terahertz pulses as a stimulus to coherently control the magnon response~\cite{kampfrath_coherent_2011,deb_2D_prb_2022}.
In this paper, we combine the concepts of cavity engineering with multiple THz pulse excitation to demonstrate the vast tuning possibilities of the coupling strength between light and magnetization at THz frequencies. Specifically, we analyze the practically important case where the cavity is created by the two interfaces of a plane-parallel micron-scale plate made of the ferrimagnet itself. The tunability of the exchange mode frequency through temperature control allows us to couple it with multiple cavity modes in a controllable way. By performing THz-pump optical-probe experiments, we monitor the amplitude and the frequency of the exchange mode to reveal the corresponding magnon-photon coupling strength at various temperatures. The enhancement of the mode amplitude is readily observable when the frequency of the magnon mode is resonant with an eigenmode of the cavity, while a reduction of the effective damping is observed across the full temperature range between 80 and 130 K. Numerical simulation based on Landau-Lifschitz-Gilbert equations in a two-sublattice system in a cavity show good agreement with the experimental findings. Overall, this work touches on a previously unexplored realm of THz cavity spin dynamics and provides a systematic understanding of the dynamical magnetic response of a ferrimagnetic cavity.

\section{MATERIALS AND METHODS}
A 200~$\mu$m thick single crystal of (Gd,Bi)$_3$Fe$_5$O$_{12}$ (GdBIG) oriented along the (111) direction was grown using liquid phase epitaxy and polished to the prescribed thickness~\cite{parchenko_non-thermal_2016}.
The sample was mounted in a continuous flow cryostat and cooled using liquid nitrogen with a dc magnetic field applied in the sample plane. Broadband THz pulses were generated using the tilted-pulse-front optical rectification technique in a LiNbO$_3$ crystal~\cite{hirori_single-cycle_2011}.
For that optical pulses from a Ti:sapphire amplifier with 6~mJ per pulse, 1~kHz repetition rate and a center wavelength of 800~nm were used.
Part off this beam was split of to be used as a low intensity probe pulse.
Two types of THz spectroscopy experiments were conducted: THz time-domain (TDS) spectroscopy and THz-pump optical-probe (TPOP).
In TDS, the THz beam was transmitted through the GdBIG sample, after which the THz pulse field strength was detected via electro-optic sampling in a 1~mm thick ZnTe crystal~\cite{wu_freespace_1995}.
TPOP was performed by overlapping the 800~nm probe beam with the THz beam within the sample.
The THz-induced changes of the probe polarization were detected  using the balanced photodiode technique~\cite{wu_freespace_1995}.
Both the THz and optical radiation were at normal incidence.
For both experimental configurations, the polarization state and fluence of the THz radiation was controlled using two wire grid polarizers.
In TDS the THz electric field strength was reduced below 10~$\frac{\text{kV}}{\text{cm}}$, whereas for TPOP, THz field strengths up to 700~$\frac{\text{kV}}{\text{cm}}$ were used.

\section{RESULTS}

\begin{figure}[t]
\centering
\includegraphics[width=8.636cm]{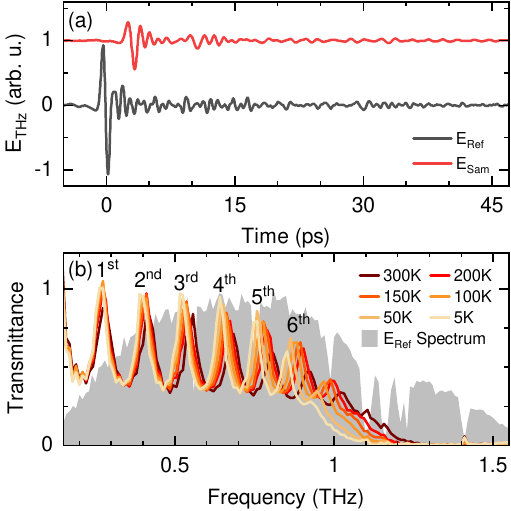}
\caption[Static]{Static characterization of GdBIG. (a) THz electric field transmitted through the sample $E_{\text{Sam}}$ and an empty sample holder $E_{\text{Ref}}$ at room temperature, offset for better visibility. (b) Transmission spectrum of GdBIG at different temperatures. The reference spectrum is shown as a gray background, where the sharp absorption lines result from the water vapor absorption. The cavity modes are numbered according to their order.}
\label{fig:1}
\end{figure}

\begin{figure}[t]
\centering
\includegraphics[width=8.636cm]{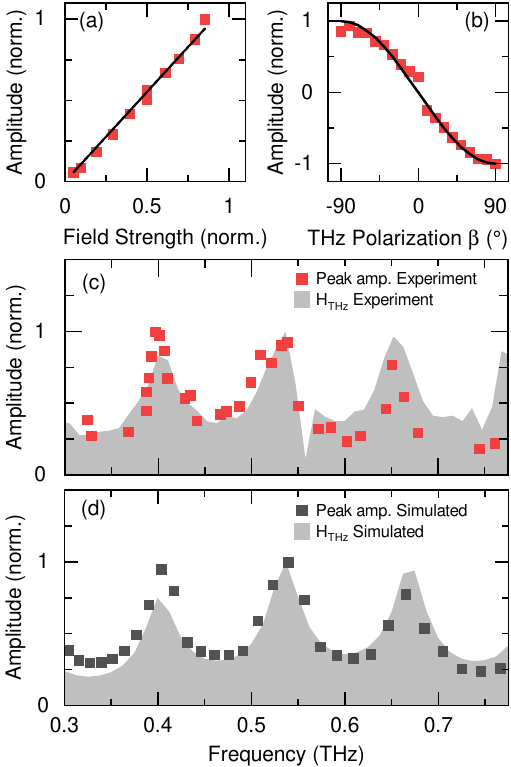}
\caption[Zeeman excitation mechanism]{THz light - exchange mode coupling mechanism. The exchange mode amplitude as a function of THz pulse field strength (a) and THz field polarization (b). These measurements were performed at a temperature of 100 K. The solid lines are the expected dependencies for a Zeeman excitation mechanism. (c) Amplitude of the exchange mode as function of its frequency. The spectrum of the transmitted pulse train corresponding to the magnetic field of the THz pulse $H_{\text{THz}}$ is shown as a gray background. (d) Amplitude of the exchange mode as function of its frequency derived from the simulations (see Sec.~\ref{sec:theory}). The simulated excitation spectrum of $H_{\text{THz}}$ is shown as a gray background.}
\label{fig:3}
\end{figure}

As a first step, the transmission properties of GdBIG were studied by THz TDS spectroscopy at different temperatures. Figure~\ref{fig:1}(a) shows the THz electric field transmitted through the sample $E_{\text{Sam}}$ and the reference THz electric field $E_{\text{Ref}}$ at room temperature. The main peak of $E_{\text{Sam}}$ is delayed by $\sim$3.1~ps compared to the reference signal.
Furthermore, a second peak at $\sim$10.7~ps is visible, originating in the back reflection of the THz pulse from the rear interface of GdBIG. For a sample thickness of 200~$\mu$m and the time delay observed in $E_{\text{Ref}}$ the effective refractive index $n$, in the $\sim 1$~THz frequency range, is estimated to be $\sim$5.6. The amplitude transmission spectra plotted in Fig.~\ref{fig:1}(b) are calculated as the ratio of the absolute values of the $E_{\text{Sam}}$ and $E_{\text{Sam}}$ FFT spectra. A strong periodic modulation of the transmission is observed, which is caused by formation of a THz pulse train due to multiple reflections from both crystal interfaces~\cite{hecht_optics_nodate}. The peaks shift to lower frequencies with decreasing temperature indicating an increase in the refractive index. Moreover, the sample becomes non-transparent above 1.3~THz at 300~K and above 1.1~THz at 5~K, indicating the presence of an absorption line outside of the spectral window which is softening with decreasing temperature. This absorption line is absent in pure Gd$_3$Fe$_5$O$_{12}$~\cite{sievers_far_1963}, which suggests that this could be a phonon mode involving Bi in the interstitial sites. This notion is reinforced by the appearance of a Raman-active optical phonon at 1.8~THz when substituting Gd with Tb in Gd$_{2.34}$Tb$_{0.66}$Fe$_5$O$_{12}$~\cite{grunberg_optical_1971}. Without changing the coupling strength, the mass ratio between the Tb and Bi ions would lower this phonon frequency to $\approx$1.6~THz. Furthermore, room temperature Raman measurements (Fig.~\ref{fig:S2}) shows the presence of a peak at 1.28 THz. According to the mutual exclusion principle~\cite{j_michael_hollas_modern_2003}, the appearance of this mode in both Raman and IR spectra can be explained by the lowering of the crystal symmetry due to the random distribution of Bi dopants in GdBIG~\cite{kumar_SHG_2010}.\\
\begin{figure}[t]
\centering
\includegraphics[width=8.636cm]{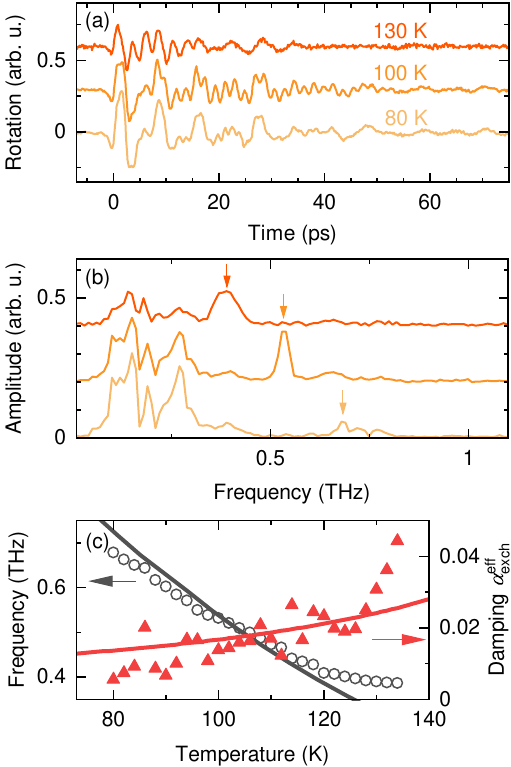}
\caption[Exchange mode]{Temperature dependence of the Gd-Fe exchange mode. (a) THz-induced polarization rotation for 80~K, 100~K, and 130~K, offset for better visibility. (b) Fourier spectra of the time traces from panel (a), the arrows indicate the exchange mode, offset for better visibility. (c) The frequency (black circles) and the effective damping (red triangles) of the exchange mode as a function of temperature. Both values are extracted with a Lorentzian function fit to the exchange mode peak. The black line shows a fit based on the Kaplan-Kittel exchange resonance formula (see Eq.~S5)~\cite{kaplan_exchange_1953} with the molecular field parameter $\text{N}_{\text{Gd-Fe}} = -0.80 \frac{\text{mol}}{\text{cm}^3}$. The red line is a fit with $\alpha^{\text{eff}}_{\text{exch}}= \alpha^{\text{eff}} \left(|\vec{M}_{\text{Gd}}| + |\vec{M}_{\text{Fe}}|\right) / \left(|\vec{M}_{\text{Gd}}| - |\vec{M}_{\text{Fe}}|\right)$~\cite{schlickeiser_damping_theory_2012} with the constant effective Gilbert damping factor of the sublattices $\alpha^{\text{eff}} = 0.0046$.}
\label{fig:2}
\end{figure}
\indent Next, TPOP spectroscopy was performed. Figure~\ref{fig:2}(a) shows THz induced polarization rotation at three distinct temperatures of 130~K, 100~K, and 80~K. The time traces indicate a rich frequency spectrum which is strongly temperature-dependent. Corresponding Fourier spectra are plotted in Fig.~\ref{fig:2}(b) showing multiple peaks composing the induced polarization rotation in Fig.~\ref{fig:2}(a). Two of the most pronounced features, centered at 0.13~THz and 0.26~THz, do not move with temperature, while the frequency of the 0.4~THz peak at 130~K increases to 0.7~THz upon lowering the temperature down to 80~K. The extracted peak position with a temperature step of 2~K is shown in Fig.~\ref{fig:2}(c), in an excellent agreement with the temperature dependence observed for the exchange resonance between the Gd and Fe sublattices \cite{stupakiewicz_squid_2021}. The positions of the other visible peaks in Fig.~\ref{fig:2}(b) that do not change with temperature within our measurement accuracy, are consistent with the 1st and the 2nd cavity modes in  Fig.~\ref{fig:1}(b). Therefore, they can convincingly be attributed to the forced response of the spin system to the THz pulse, shaped by multiple reflections at GdBIG surfaces.
Moreover, the cavity modes modify the observable line-width of the magnetic resonance, thereby enabling control of the apparent effective damping of the exchange mode, $\alpha^{\text{eff}}_{\text{exch}}$. The apparent change in damping results from the recurrent parametric excitation of the spin system by a phase-matched sequence of the THz pulses within the cavity. The extracted line-width of the exchange mode at every temperature is shown in Fig.~\ref{fig:2}(c). The value of $\alpha^{\text{eff}}_{\text{exch}}$ varies from 0.007 to 0.04 in the observed temperature range. This is almost one order of magnitude lower than previously reported measurements on similar compounds. With a non-resonant excitation the intrinsic damping of the exchange mode was shown to be from 0.05 to 0.1~\cite{deb_damping_2022}. Rigorous estimation of the enhancement is challenging due to limitations in frequency resolution and the low finesse of the cavity. However, we observed that the enhancement is consistent with the finesse, measuring 5-6 in our experiment. \\
\indent Controlling the exchange mode necessitates understanding the underlying excitation mechanism. To this end, the THz field strength dependence of the induced magnetization dynamics was analyzed. As illustrated in Fig.~\ref{fig:3}(a), the amplitude scales linearly with the strength of the THz field. Moreover, the change of exchange mode amplitude with changing THz field polarization $\beta$, where 
$\beta = \angle{(\vec{M},\vec{H}_\text{THz})}$, shown in Fig.~\ref{fig:3}(b), reveals a sine-like dependence. The two observed maxima at $\beta=90^{\circ}$ and $\beta=-90^{\circ}$ are in agreement with the linear Zeeman torque excitation mechanism~\cite{kampfrath_coherent_2011}.\\
\indent Taking advantage of the high tunability of the exchange mode with temperature, we analyzed its coupling with the cavity modes and their influence on the resulting spin dynamics. Figure~\ref{fig:3}(c) shows the amplitude of the exchange mode as a function of its frequency, where each data point corresponding to a distinct temperature. Although the net magnetic moment of GdBIG significantly decreases as the temperature approaches the compensation point $T_{\text{comp}} = 223$~K~\cite{parchenko_non-thermal_2016}, this monotonic trend does not account for the observed periodic behavior. However, consistent with the Zeeman torque excitation mechanism, the amplitude of the exchange mode linearly depends on the THz field strength. To illustrate this, we present the spectrum of the magnetic field of the transmitted THz pulse, $H_{\text{THz}} \propto E_{\text{Sam}}$, measured with EOS at 100 K, shown as a gray background in Fig.~\ref{fig:3}(c). Indeed, the exchange mode amplitude increases twofold when its frequency matches one of the cavity resonances.

\section{THEORETICAL ANALYSIS AND DISCUSSION}
\label{sec:theory}
To understand the THz-induced magnetization dynamics in the presence of cavity modes, we employ the modified Landau–Lifshitz–Gilbert equation formalism. Taking into account the dominant role of the iron-iron exchange, the magnetization dynamics can be described by coupled equations for a two-sublattice ferrimagnet~\cite{geschwind_GdIG_1959}:
\begin{align}
    \frac{d\vec{M}_{\text{i}}}{dt} &= -(\mu_0\gamma_{\text{i}}) \left[ \vec{M}_{\text{i}} \times \vec{H}^{\text{eff}}_{\text{i}}\right] - ( \mu_0\gamma_{\text{i}}) \frac{\alpha_{\text{i}}}{\lvert \vec{M}_{\text{i}}\rvert} \vec{M}_{\text{i}} \times \left[ \vec{M}_{\text{i}} \times \vec{H}^{\text{eff}}_{\text{i}} \right]
\end{align}
with $\mu_0$ the vacuum magnetic permeability, $M_\text{i}$ the net magnetizations, $\gamma_\text{i}$ the gyromagnetic ratios, and $\alpha_\text{i}$ the Gilbert damping factors of the Gd and Fe sublattice dynamics. $\vec{H}^{\text{eff}}_{\text{i}}$ is the effective magnetic field defined as
\begin{align}
\label{eq:effective_field}
    \vec{H}^{\text{eff}}_{\text{Gd,Fe}} = \vec{H}_{\text{app}} + \vec{H}_{\text{THz}} -  N_{\text{Gd-Fe}}\vec{M}_\text{Fe,Gd}.  
\end{align}
Here, the first term with $\mu_0\vec{H}_\text{app} = 150$~mT represents the applied external magnetic field, $\vec{H}_\text{THz}$ denotes the THz pulse magnetic field, and $N_{\text{Gd-Fe}}$ the effective Gd-Fe exchange parameter. To model the impact of the cavity, we introduce multiple delayed THz pulses so that the temporal profile of the THz stimulus closely follows the experimentally observed transmitted field $E_{\text{Sam}}$ in Fig.~\ref{fig:1}(b) at 100~K. A comparison of the simulated and measured THz waveforms and spectra is presented in Fig.~\ref{fig:S1}(a,b) of the Appendix B. 

The exchange field responsible for the coupling of the Gd and joint Fe sublattices is represented by the last term in Eq.~\ref{eq:effective_field}. It is worth noting that the two-sublattice model can be obtained as a limiting case of the full three-sublattice description, where the Fe-Fe exchange interaction is infinitely stronger than the Gd-Fe one. This three-sublattice formalism is bulky yet instructive, relating an effective $N_{\text{Gd-Fe}}$ parameter to the parent molecular field coefficients $N_{\text{cd}}$ and $N_{\text{ca}}$. Moreover, it is capable of accounting for the temperature dependence of the effective Gd-Fe coupling, which can be highly relevant in our case. Our analysis (see Appendix C),  however, indicates very small ($\sim1\%$) variations of $N_{\text{Gd-Fe}}$ in the considered temperature range. With that in mind, we retain the two-sublattice approach to fit the exchange mode frequency with the Kaplan-Kittel resonance (see Fig.~\ref{fig:2}(c)). We obtain $N_{\text{Gd-Fe}}$ = -0.8~$\frac{\text{mol}}{\text{cm}^3}$, in good agreement with the value derived from the three-sublattice model (see Appendix C). Far away from the compensation point $T_{\text{comp}} = 223$~K, it is safe to assume that the Gilbert damping factors are equal $\alpha_{\text{Fe}}=\alpha_{\text{Gd}}=\alpha$~\cite{schlickeiser_damping_theory_2012} and almost temperature independent~\cite{deb_damping_2022}, while their values were set to 0.03 based on the previous measurements on this very sample~\cite{parchenko_non-thermal_2016}. The gyromagnetic ratios are assumed equal as well $\gamma_{\text{Fe}} \approx \gamma_{\text{Gd}} = \gamma$ = $28 \frac{\text{GHz}}{\text{T}}$~\cite{geschwind_GdIG_1959}. 

By numerically solving Eqs. (1) using the Euler method, the magnetization dynamics of the sublattices was extracted. To compare the simulation results with the experiment it is important to note that the Faraday rotation at 800~nm, in our experimental geometry, is mostly sensitive to the magnetization of the iron sublattice in GdBIG~\cite{parchenko_non-thermal_2016}. Figure~\ref{fig:S1}(c) shows the simulated and experimentally observed THz-induced magnetization dynamics at 100~K. Both time scans show a fast oscillation attributed to the exchange mode dynamics. The experimental scan has an additional low frequency component  that is not captured by our simulations. However, the simulation does not take into account propagation effects and dispersion of the sample, therefore an exact match is not expected. Simulating temperature changes by varying the sublattice magnetizations allows us to extract the amplitude and the frequency of the exchange mode shown in Fig.~\ref{fig:3}(d). The gray background illustrates the spectrum of the applied THz magnetic field $H_\text{THz}$. Similar to the experiment, the amplitude of the exchange mode is enhanced when its frequency coincides with the cavity modes created by the sample interfaces.\\
\indent Note that in the case of opto-magnetic excitation~\cite{parchenko_wide_2013} of the exchange mode, much weaker recurrent magnon pumping can be anticipated. Two main factors responsible for this reduction can be outlined. First, the optical refractive index is significantly smaller than that in the THz range~\cite{hibiya_ref_index_1985}, resulting in weaker reflected pulses. Second, an effective opto-magnetic field which is instrumental for the excitation, is proportional to the square of the electric field of light. Because of this, the effective magnetic field generated by the second and subsequent optical pulses in the cavity will be drastically reduced.\\
\indent We believe that the concept of driving high-frequency spin dynamics with a THz pulse train in a cavity formed by the crystal interfaces can be further explored. For example, we suggest a sandwich structure consisting of a THz generation crystal and a ferrimagnet capped by optically transparent and highly THz reflective interfaces. The progress in THz coating makes it possible to achieve very high reflection coefficients, which accentuate the action of the THz pulse train on the medium. For example, indium–tin–oxide-coated glass has THz amplitude reflection coefficient $r$ as high as 0.9~\cite{ito_bauer_2002}, while using THz Bragg mirrors this value potentially can reach 0.99~\cite{bragg_yu_2018}. This is substantially higher than that obtained in a bare GdBIG crystal with Fresnel reflection at interfaces, where we estimate $r$ = 0.7. To study the impact of interface reflection, we simulate THz-induced magnetization dynamics with $r$ = 0.95 and plot the amplitude as a function of the frequency of the exchange mode in Fig.~\ref{fig:S3}(a). A much higher dynamical range of the mode amplitude variations is observed, while $\alpha^{\text{eff}}_{\text{exch}}$ can be further reduced by fourfold. This is illustrated in Fig.~\ref{fig:S3}(b)  by comparing magnetization dynamics waveforms at a temperature of 128~K. Moreover, in conjunction with using ferrimagnets with much lower intrinsic damping, e.g., yttrium iron garnet~\cite{CHEREPANOV_YIG}, exchange mode coupling with THz light might even enter the strongly coupled regime ~\cite{strong_coupling_bialek,strong_coupling_huebl}.

\section{CONCLUSION}
In conclusion, we investigated the coupling between light at THz frequencies and the Gd-Fe spin exchange mode in GdBIG using THz-pump optical-probe measurements. We show that tuning the exchange mode into a resonance with the cavity modes strongly enhances the efficiency of the THz excitation of high-frequency spin dynamics. The excitation mechanism is attributed to the Zeeman torque. Moreover, the cavity reduces the effective damping of the exchange mode 5-fould. The results are analyzed by means of numerical simulations based on a system of coupled Landau-Lifshitz-Gilbert equations which show a good agreement with the experimental observations. A three-subsystem approach developed here for the first time allowed to conclude that the effective Gd-Fe molecular field strength exhibits no variations in the 70-130~K temperature range. This work improves the understanding of efficient control of the exchange modes at THz frequencies. Furthermore, it opens an intuitively clear and uncomplicated path to achieve strong coupling between light and sub-THz exchange modes by means of designing the effictive interfaces of a magnetic material.\\

The supporting data and codes for this article are available from Zenodo~\cite{gdbig_data}.\\

\section*{ACKNOWLEDGMENTS}
We thank T. Satoh for providing the samples and fruitful discussions.
The work in Cologne was partially supported by the DFG via Project No. 277146847 — Collaborative research Center 1238: Control and Dynamics of Quantum Materials (Subproject No. B05).

\section{Appendix}

\renewcommand{\thefigure}{S\arabic{figure}}
\renewcommand{\theequation}{S\arabic{equation}}
\setcounter{figure}{0}
\setcounter{equation}{0}

\subsection{Raman Measurement}

\begin{figure}[htb]
\centering
\includegraphics[width=8.636cm]{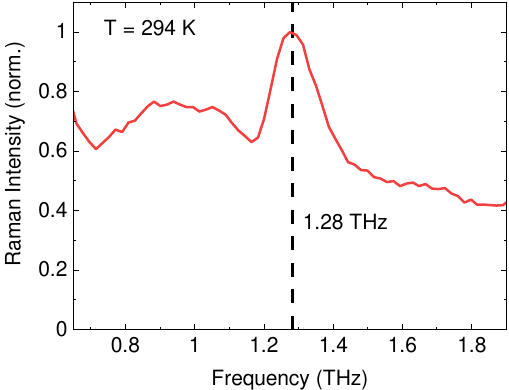}
\caption[Raman spectrum]{Raman intensity measured on GdBIG in parallel polarization configuration at room temperature.}
\label{fig:S2}
\end{figure}

The Raman spectrum was recorded by a triple-grating spectrometer using a 532~nm continuous wave laser in backscattering geometry at a power of 1~mW on the sample. The direction of polarisation of the analysed light was parallel to that of the incoming beam. 
The Raman spectrum in the relevant spectral range is shown in Fig.~\ref{fig:S2}.

\subsection{Modeling of THz Magnetic Field}

\begin{figure}[htb]
\centering
\includegraphics[width=8.636cm]{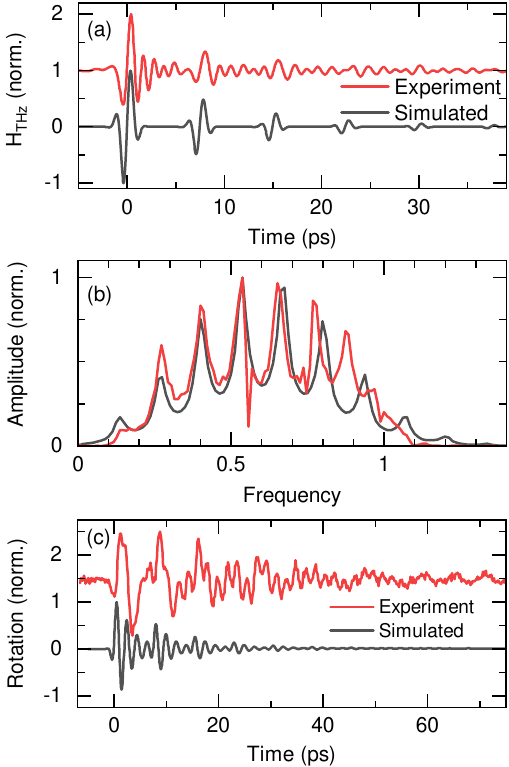}
\caption[Spectrum comparison]{(a) Magnetic THz field applied in the simulation compared to the experimentally measured THz field transmitted through the sample at 100~K. (b) Corresponding Fourier spectra. (c) Comparison of the THz-induced magnetization dynamics in the sample between experiment and simulation at 100~K.}
\label{fig:S1}
\end{figure}

\begin{figure}[htb]
\centering
\includegraphics[width=8.636cm]{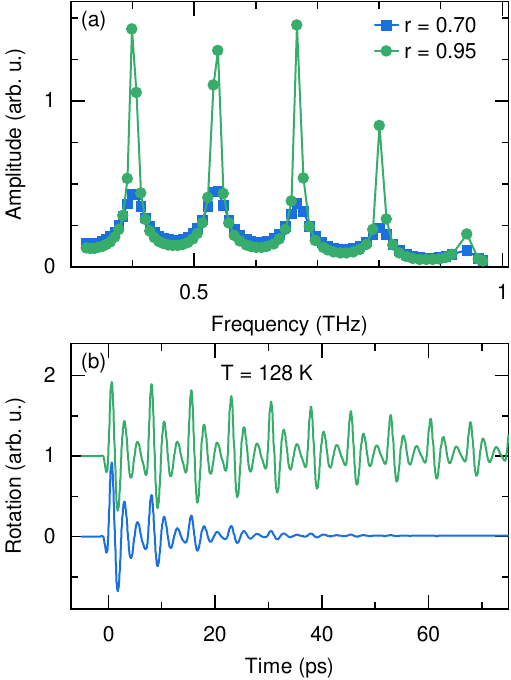}
\caption[Spectrum comparison]{(a) Numerical simulation of the behavior of the exchange mode amplitude as a function of its frequency with two different reflection coefficients at the interface. (b) THz-induced magnetization dynamics modeled with the reflection coefficient $r = 0.7$ (bottom curve) and $r = 0.95$ (top curve). }
\label{fig:S3}
\end{figure}

To model the THz pulse magnetic field $H_{\text{THz}}$, we used sequences of pulses defined as

\begin{align}
    H_{\text{THz}}(t) = H_{0}\sum_{k=0}^N \cdot r^{2k} \text{sin}(\omega_0t-k\cdot \omega_0\tau) \cdot e^{-(t-k\cdot \tau)^2/2\sigma^2},
\end{align}

with $H_{0}$ the magnetic field amplitude, $\omega_0$ the center frequency, $\sigma$ the pulse duration, $N$ the number of reflections, $r$ amplitude reflection coefficient at the crystal interface, and $\tau$ the time delay introduced by propagation in the medium. To follow the shape of the THz pulse used in the experiment we choose $\mu_0 H_{0} = 0.266$~T,  $\omega_0/2\pi = 0.6$~THz, and $\sigma = 0.63 $~ps. The coefficients $r = (n-1)/(n+1)$ and $\tau = 2nL/c_0$ are calculated using a constant refractive index of $n = 5.6$ extracted from the THz-TDS measurements in Fig.~\ref{fig:1}, where $c_0$ is the speed of light in vacuum and the crystal thickness $L$ = 200~$\mu$m. $N = 10$ is selected to ensure that all reflections within the chosen time window are accounted for. Figure~\ref{fig:S1}(a) shows the resulting sequences of THz pulses compared to the THz electric field measured with EOS after transmitting through the sample at 100~K. A very good agreement of the corresponding spectra is also evident, see Fig.~\ref{fig:S1}(b). To compare the THz-induced magnetization dynamics with that observed in the experiment we plot the out-of-plane component of $\vec{M}_\text{Fe}$ in Fig.~\ref{fig:S1}(c).

This approach allows for the testing of different paths to optimize the excitation efficiency of the exchange mode. Figure~\ref{fig:S3} shows the simulation results of THz-induced spin dynamics with the reflection coefficients at the interfaces of $r$ = 0.95. The other parameters are kept the same.
It is seen that improving the cavity finesse up to 60  results in a threefold enhancement of the dynamic range of the spin exchange mode peak amplitude (green circles).

\newpage
\subsection{Effective exchange constant}
Rare earth iron garnets are typically considered two-sublattice magnets. This assumption is based on the fact that the molecular field coefficient $N_\text{ad}$, which accounts for the interaction between iron at the octahedral and tetrahedral sites, is much larger than the iron-rare earth (rare earth is indexed by "$c$") molecular field coefficients $N_\text{cd}$ and $N_\text{ca}$~\cite{dionne_mol_field_1971}. For that reason, the majority of recent papers do not address the connection between the effective coupling constant $N_{\text{Gd-Fe}}$ and the three-sublattice molecular field coefficients. However, this assumption potentially overlooks the temperature dependence of the effective coupling constant~\cite{clark_3sublattice_1968}. Considering a three-sublattice model instead provides an additional way to estimate $N_\text{Gd-Fe}$ by analysing SQUID data in combination with molecular field theory. The main steps for deriving $N_\text{Gd-Fe}$ are provided below. 

\newpage
To address the dynamics of a three-sublattice magnet, we write three Landau–Lifshitz equations assuming no damping for simplicity:

\begin{equation*}
      \frac{d\vec{M}_{\text{a}}}{dt} = -(\mu_0\gamma_{\text{Fe}}) \left[\vec{M}_{\text{a}} \times \left(\vec{H}_{\text{THz}}-N_\text{ad} \vec{M}_{\text{d}}-N_\text{ac} \vec{M}_{\text{c}}  \right) \right] 
\end{equation*}

\begin{equation}
    \frac{d\vec{M}_{\text{d}}}{dt} = -(\mu_0\gamma_{\text{Fe}}) \left[ \vec{M}_{\text{d}} \times \left(\vec{H}_{\text{THz}}-N_\text{da} \vec{M}_{\text{a}}-N_\text{dc} \vec{M}_{\text{c}}  \right)\right]
\end{equation}
\begin{equation*}
  \frac{d\vec{M}_{\text{c}}}{dt} = -(\mu_0\gamma_{\text{Gd}}) \left[ \vec{M}_{\text{c}} \times \left(\vec{H}_{\text{THz}}-N_\text{ca} \vec{M}_{\text{a}}-N_\text{cd} \vec{M}_{\text{d}}  \right) \right]
\end{equation*}

Then, we look for a solution in the form $\vec{M}_i = \vec{M}_i^\text{init} +\vec{m}_ie^{i 2\pi f t}$, where $\vec{M}_i^\text{init}$ are the ground state magnetizations, $\vec{m}_i$ are the corresponding perturbations, and $i$ corresponds to the indices $a$, $d$ and $c$. We assume that all magnetizations are collinearly aligned in a ground state, note that $N_\text{ad}=N_\text{da}$, $N_\text{ca}=N_\text{ac}$, and $N_\text{cd}=N_\text{dc}$, take into account terms only linear on $\vec{m}_i$ and exclude terms of the second order on  $N_\text{ac}$ and $N_\text{dc}$. Then, the Gd-Fe exchange frequency can be found:
 \begin{widetext}
\begin{multline}
f_{\text{Gd-Fe}}=\frac{\mu_0\gamma}{2} \left[ 
N_\text{ad} \left( M_d - M_a \right)
- N_\text{ac} \left( M_a + M_c \right)
+N_\text{dc} \left( M_d - M_c \right)\right]-
\\ \frac{\mu_0\gamma}{2}\left[ 
N_\text{ad}^2 \left( M_a^2 + M_d^2 - 2 M_a M_d \right)
- 2 N_\text{ac} N_\text{ad} \left( M_a^2 + M_a M_c - M_a M_d +
M_c M_d \right)
- 2 N_\text{ad} N_\text{dc} \left( M_d^2 - M_a M_d - M_a M_c - M_c M_d \right)\right]^{1/2}
\end{multline}
\begin{multline}
\label{eq:I_dont_care}
f_{\text{Gd-Fe}}=\frac{\mu_0\gamma}{2} \left[ 
N_\text{ad} \left( M_d - M_a \right)
- N_\text{ac} \left( M_a + M_c \right)
+N_\text{dc} \left( M_d - M_c \right)\right]-
\\ \frac{\mu_0\gamma}{2} N_\text{ad} \left( M_d - M_a \right)\left[1- 2 \left(\frac{N_\text{ac}}{N_\text{ad}}\right) \frac{M_a^2 + M_a M_c - M_a M_d +
M_c M_d}{M_a^2 + M_d^2 - 2 M_a M_d}
- 2 \left(\frac{N_\text{dc}}{N_\text{ad}}\right)\frac{M_d^2 - M_a M_d - M_a M_c - M_c M_d}{M_a^2 + M_d^2 - 2 M_a M_d}\right]^{1/2}
\end{multline}
\end{widetext} 

Given that the iron-iron exchange interaction considerably exceeds the rare earth-iron exchange interaction, $N_\text{ac}$, $N_\text{dc} << N_\text{ad}$, the last two terms inside the square root in Eq.~\ref{eq:I_dont_care} represent a small parameter to which a Taylor expansion can be applied.  This results in 
\begin{align}
  f_{\text{Gd-Fe}}=-(\mu_0\gamma) N_{\text{Gd-Fe}}\left|M_\text{d}-M_\text{a}-M_\text{c}\right|;\\
  \label{eq:exchange}
  N_{\text{Gd-Fe}}=-{\frac{N_\text{cd}M_\text{d}+N_\text{ca}M_\text{a}}{M_\text{d}-M_\text{a}}}.
\end{align}

Equation~\ref{eq:exchange} clearly shows that since the magnetizations $\vec{M}_{\text{d}}$ and $\vec{M}_{\text{a}}$ are temperature-dependent, the effective Gd-Fe exchange coupling also exhibits temperature dependence when the three-sublattice model is considered. Substituting molecular field coefficients $N_\text{cd}$ =  6~$\frac{\text{mol}}{\text{cm}^3}$ and $N_\text{ca}$ = -3.44~$\frac{\text{mol}}{\text{cm}^3}$ from~\cite{dionne_mol_field_1971} and deriving $M_{\text{d}}$, $M_{\text{a}}$ and $M_{\text{c}}$ from SQUID data~\cite{parchenko_laser-induced_2016} give an estimation of $N_\text{Gd-Fe}$ changing in the range from -0.89 to -0.90~$\frac{\text{mol}}{\text{cm}^3}$ in the studied temperature range from 80 to 130~K.

\bibliographystyle{apsrev4-2}
\bibliography{GdBIG_bib.bib}
\end{document}